\documentclass[12pt, preprint]{aastex}
\usepackage{emulateapj5}  

\def\einstein{{\it Einstein }}	
\def\ergcm2s{~erg cm$^{-2}$ s$^{-1}$ } 
\def\ergs{~erg s$^{-1}$}		
\def\lunit{~erg s$^{-1}$}		
\def\cmsq{~cm$^{-2}$ }		%
\def\nh{~$\rm{N_{H}}$}

\def\lx{~$\rm{L_{X}}$}
\def\etal{et al.~}		
\def\kms{~km s$^{-1}$}		
\def\kmsmpc{~km s$^{-1}$ Mpc$^{-1}$}
\def\msun{~M$_{\odot}$}
\def\deg{$^{\circ}$}
\def\n4038{~NGC4038/39}		
\def\chandra{{\it Chandra }}
\def\sherpa{{\it Sherpa }}
\def\x2{$\chi^{2}$}	



\begin{document}

\title{The X-ray Luminosity Function of ``The Antennae'' 
Galaxies (NGC4038/39) and the Nature of Ultra-Luminous X-ray Sources \\}

\author{ A. Zezas, G. Fabbiano}
\affil{Harvard-Smithsonian Center for Astrophysics,\\ 60 Garden
Street, Cambridge, MA 02138}
\shorttitle{\chandra\ The ULXs of ``The Antennae'' Galaxies}
\shortauthors{Zezas \& Fabbiano}
\bigskip

\begin{abstract}

We derive the X-ray luminosity function (XLF) of the X-ray source
population detected in the \chandra observation of NGC4038/39 (the Antennae).
 We explicitly include photon counting and spectral parameter uncertainties in
our calculations. The cumulative XLF is well
represented by a flat power law ($\alpha=-0.47$), similar to those describing
the XLFs of other star-forming systems (e.g. M82, the disk of M81), but different
from those of early type galaxies. This result associates  the X-ray
source population in the Antennae with young  High Mass X-ray
Binaries.
In comparison with less
actively star-forming galaxies, the XLF of the Antennae has a highly
significant excess of sources with luminosities above $10^{39}$\ergs~
(Ultra Luminous Sources; ULXs).
We discuss the nature of these sources, based on the XLF and on their
general spectral properties, as well as their optical counterparts
discussed in Paper~III. We conclude that  the majority of the ULXs
 cannot be intermediate
mass black-holes (M$ > 10-1000$\msun) binaries, unless they are linked to the 
remnants of massive Population III stars (the Madau \& Rees model). 
Instead, their spatial
and multiwavelength properties can be well explained by beamed emission
as a consequence of supercritical accretion. Binaries with a neutron
star or moderate mass black-hole (up to 20\msun), and B2 to A type star 
companions would be consistent with our data.
In the beaming scenario, the XLF should exibit caracteristic breaks that
will be visible in future deeper observations of the Antennae.

\end{abstract}


\section{Introduction}

 It is well known that the  Galaxy hosts a population of X-ray sources
 associated with X-ray binaries (XRBs) and supernova remnants (SNRs)
(e.g. review in Watson \etal 1990; Grimm \etal 2001).  Observations with the \einstein
Observatory and later with ROSAT and ASCA, detected individual X-ray sources in nearby 
galaxies  and initiated the study of the
X-ray source populations in galaxies (e.g. Fabbiano 1989, Fabbiano
1995,  Read \etal 1997; Wang
\etal 1999; Yokogawa \etal 2000). 
X-ray Luminosity Functions (XLFs) were 
derived for a few  galaxies (e.g. M31, M81, M101;  Trinchieri  \& Fabbiano 1991, Fabbiano 1988, Fabbiano 1995,
Primini \etal 1993, Supper \etal 2001, Wang \etal 1999), but these studies 
were of necessity limited by the available observational capabilities. Even 
so, it was clear that using samples of X-ray sources in external galaxies 
would reduce the problems in the study of the population properties of 
Galactic sources, inherent to our position in the Galaxy (Fabbiano 1995). 
Moreover, the XLFs reflect the formation, evolution, and physical properties 
of the X-ray source population in a galaxy, so comparison of XLFs of different 
galaxies, and modeling of the same, provide powerful tools for understanding 
the nature of the X-ray sources and relating them to the evolution of the 
parent galaxy and its stellar population.

With {\it Chandra} (Weisskopf et al 2000), X-ray source population studies are 
finally coming of age. The sub-arcsecond resolution of the {\it Chandra} 
mirrors (Van Speybroeck et al 1997) allows both the separation of discrete 
sources from surrounding diffuse emission and the detection of much fainter 
sources than previously possible. The first {\it Chandra} studies of XLFs 
suggest trends related to the morphological type of the parent galaxy and/or 
the age of the prevalent stellar population: XLFs of early-type galaxies tend 
to have a break at the Eddington Luminosity of neutron star XRBs
(e.g. Sarazin \etal 2001, Irwin \etal 2001, Blanton \etal 2001), 
while in the disk of M81, the XLF follows a single power-law
(Tennant et al 2001).  Similarily the XLFs of star-forming galaxies
can also be modeled with a single power-law (e.g.  Zezas \etal 2001;
Bauer \etal 2001; Kilgard \etal 2001)

In this paper we report the study of the XLF of the X-ray sources detected in 
the Chandra observation of NGC~4038/39, the Antennae galaxies (OBSID 315; Fabbiano et al 
2001, Paper~I; Zezas et al 2001a and b, Papers~II and III). 
Being a merging system of galaxies, the Antennae are undergoing exceptionally
vigorous star formation (e.g. Whitmore \etal 1999), and thus provide a unique laboratory for
the study of the X-ray source population in young and intense starbursts. 
The X-ray source population of the 
Antennae is exceptional, in that of the 43 point-like sources detected, 17 
have X-ray luminosity significantly in excess of that expected from normal 
neutron star or stellar-mass Black Hole (BH) XRBs.
These Ultra Luminous X-ray sources (ULXs) have spectra consistent with ULXs 
detected in other, more nearby galaxies (Paper~III), but their 
number is exceptionally large.

Here we first derive the XLFs of the Antennae and discuss uncertainties and
biases in its derivation (\S 2), and we  compare it with the 
{\it Chandra} XLFs of other galaxies (\S 3). We then use these results, in 
conjunction with the spectral properties of the X-ray sources, and the 
information on their optical counterparts presented in Paper~III, to constrain 
models for the ULX population.
Although the luminosities in Paper~II were derived for $\rm{H_{\circ}=50}$~\kmsmpc, we use both 
$\rm{H_{\circ}=50}$~\kmsmpc~ and $\rm{H_{\circ}=75}$~\kmsmpc~ for the present discussion.

\section{X-ray Luminosity function  (XLF) of point sources}

\subsection{Derivation of the XLF}

In order to derive the XLF of the point sources in the Antennae we
used the source 
list presented in Table 1 of Paper II, which contained
 49 sources down to a limiting luminosity of
$\sim10^{38}$\ergs~ ($\rm{H_{\circ}=50}$~\kmsmpc;
$\sim5\times10^{37}$\ergs, for  
$\rm{H_{\circ}=75}$~\kmsmpc). Of these, 31 (38, for
$\rm{H_{\circ}=75}$~\kmsmpc) have luminosities below
$10^{39}$\ergs~ 
while 18 (11) have higher luminosities.  Forty-three sources appear
point-like; of these 26 (33)  have luminosities below
$10^{39}$\ergs~  and 17 (10) have higher luminosities.  The luminosities were
calculated for a 5~keV bremsstrahlung model and  Galactic column
density ($\rm{N_{H}=3.24\times10^{20}}$\cmsq; Stark \etal 1992) and are corrected for absorption. Fig.~1 shows the luminosity
distribution of these sources.   The calculation of an
unbiased XLF from this luminosity distribution is not straightforward because of the  spatial
variations of the intensity of the diffuse 
emission in the Antennae galaxies (see Paper~I; Fabbiano \etal 2001)
 which cause  different detection thresholds over the
system.
 To account for this effect without setting an overly
conservative uniform completeness limit, 
 and thus narrowing the luminosity
coverage, we calculated a spatially-dependent correction function
(CF, see \S2.2), and corrected the XLF accordingly. This CF estimates the number of sources 
which escaped detection  as a function of the source
 luminosity and the background surface brightness.

 Given the limited source statistics,  we calculated  
 XLFs for sources detected in regions with just three background levels:
0.0 - 0.15~ct/pix (0.0 - 1.2~ct per typical detection cell; Paper~II),
0.15 - 0.35~ct/pix (1.2 - 2.8~ct per detection cell) and
 greater than 0.35~ct/pix (2.8~ct per detection cell). 
These background levels were determined
from the background map produced by the CIAO {\textit{wavedetect}}  tool, which is  a
reconstruction of the background based on the wings of the Mexican Hat
function after subtracting all the point sources (Dobrzycki \etal 2000). 
The background levels were chosen in order to have at least 8 sources in each XLF.
Fig. 2 shows the detected sources on the background map of the
Antennae. On the same map we present as contours background levels of
0.01~cnts/pix, 0.15~cnts/pix and 0.35~cnts/pix.  We also mark the
detected sources following the notation of Table~1 from Paper~II. 
In the calculation of the XLFs we did not include the
 extended sources (sources 5, 6,
7, 10, 24 and 29) and the Southern nucleus (the Northern nucleus is
excluded as an extended source). 

We also excluded 
the three hard sources which were detected in the
obscured contact region between the two galaxies (sources 35, 39 and 40),
 because of the difficulty in correcting the LF for the
effect of foreground obscuration which varies from source to source.
The obscuration correction is further complicated by our ignorance of
the  three dimensional distribution of the 
X-ray sources and  the obscuring material. The latter is very hard
to determine from the spectra of the sources, since it is unknown
what fraction of the absorption is due to foreground material and
what is due to intrinsic absorption in the source.  
 The exclusion of these sources will not bias our XLF as the excluded area is very small compared to
the total area of the Antennae. The hatched histogram in Fig~1 shows
the luminosity distribution of 
the sources used for the determination of the XLF.  The three uncorrected XLFs are shown
in Fig. 3a.

\subsection{The Correction Function}

 In order to determine the CF we simulated 50 realizations of 
 three different datasets, one for each background level,
with sources of luminosities in the range $10^{37}-10^{39}$\lunit.
For these simulations we 
used the same spectrum as was used to calculate
 the luminosity of the sources (see \S2.1).  For the spectrum of the
background we assumed a flat power-law  as determined from spectral
fits in  a source free area of the ACIS-S3 chip.
We ran {\textit{wavedetect}} on each simulated dataset with exactly
the same parameters as used for the source detection in the actual
data. 
The 50 realizations give an uncertainty of 15-20\% in the CF
 between the 100\% and the 50\% completness limit, which is smaller
than the Poisson error from the source counts.

The resulting CF is presented in Fig.~3b. The square points correspond to low background (0.1~cnts/pixel),
the circles correspond to medium background (0.25~cnts/pixel)
and the stars correspond to high background (0.45~cnts/pixel).
The correction is important only for the lowest luminosity
sources ($\rm{L_{X}<1.6\times10^{38}}$).
The three curves differ close to the
detection limit, the 50\%  completeness limit (indicated by a dashed
line) ranges between
$8\times10^{37}$\lunit and $1.3\times10^{38}$\lunit for all
background levels.  Since there
are only 5 sources in this luminosity range, we can conservatively set our
completeness limit at $1.3\times10^{38}$\lunit.

 The convolution of the three XLFs (presented in Fig.~3a) with the
appropriate CF for each background (Fig.~3b) provides the completeness corrected
XLF for each background level. This correction is important only
in the luminosity range between the detection limit and
$2\times10^{38}$\ergs, where there are 7, 2 and 1 sources in the high,
medium and low background XLFs respectively. Above this luminosity no
correction needs to be applied (Fig.~3b). The corrected XLF is
shown by the dashed line in Fig.~3a.
 Adding together the corrected 
 XLFs gives the total XLF of the 
Antennae galaxies.  The top left panel of Fig.~4 shows
the differential total XLF whereas the top right  panel shows
the cumulative XLF.
 The errors in both plots are calculated according to the
Gehrels statistic \footnote{A modification of Gaussian statistics to
approximate a Poisson distribution. The standard deviation of the
Gaussian is  $\rm{\sigma = 1 + \sqrt(0.75 + N)}$.}
(Gehrels 1986) and do not include uncertainties in 
the correction function.
 This XLF is complete down to
$1.3\times10^{38}$\ergs~ which is the 50\% completeness limit 
in the high background case. The application of the correction
function and the  50\% completeness will minimize the effect of the
Eddington bias.
 Because of the applied corrections we cannot use the unbinned XLF 
for any further analysis.
  
\subsection{Uncertainties in the determination of the XLF}
 
 The XLF as calculated above includes only errors associated with the Poissonian nature 
of the source detection process (source counting uncertainties). However, an XLF is also subject to 
 uncertainties related to the determination of the luminosities for
each source. These are:
 (i) the uncertainties in the count-rate of each source due to the Poissonian nature of the photon
 detection for each source (photon counting uncertainties) and (ii) uncertainties in the determination 
of the spectral shape of each source (spectral uncertainties).
In order to determine the effect of  these additional sources of error,  we performed a Monte-Carlo
simulation to obtain 1000 estimates of the count rate for each source. Although the photon detection process is governed 
by Poisson statistics, the use of Gaussian statistics is a fairly good approximation when dealing with sources with 
 more than $\sim20$ counts (e.g. Bevington \& Robinson 1992). In this case the most appropriate determination
 of the standard deviation is given by the Gehrels formula (Gehrels 1986). Hence, in these simulations
 we used Gaussian deviates  with mean equal to the number of detected counts and standard deviation given 
by the Gehrels statistic. The errors in the XLF due to photon
counting uncertainties range from $\pm0.2$ for the high luminosity
bins up to $\pm2.0$ sources in low luminosities.

From the simulated number of counts for each source we calculated the corresponding
luminosity, corrected only for the Galactic \nh, 
for the best fit power-law model of the coadded spectra
in the  luminosity 
range of each source  (presented in Paper~III). 
In order to obtain an estimate of
the uncertainties in the luminosity  due to spectral uncertainties we also estimated 
luminosities for the above spectra, modified by the typical
uncertainties in the photon 
indices and column densities for sources in each luminosity range given
 in Paper~III: for fainter sources we used a steeper power-law with a
larger range in the photon index  
and absorption.
The  spectral parameters  used for these estimates are given in
Table~1.  To calculate the
minimum luminosity we used the steepest slope with the highest
absorption, whereas to calculate the maximum luminosity we used the
flattest slope with the minimum absorption. The difference between the
minimum and the maximum luminosity estimates for the same count rate
is up to a factor of 2. This gives only the
extreme values of the luminosity and may overestimate the
uncertainties.  A more accurate treatment should
take into account the fact that the spectral uncertainties are
 also normally distributed and correlated, and perform a Monte Carlo simulation for
the spectral parameters. However, this is quite complicated since the
conversion factor from count rates to luminosity is not a linear
function of the spectral parameters.  The combined uncertainties due to photon
statistics and spectral uncertainties range between 0.5 and 20.0
sources ($\sim50$\%). 

 From the results of this simulation we derived the differential XLF by calculating
 the mean number of sources and the standard deviation for each 
luminosity bin using the same binning as for the observed LF presented
in Fig.~4. 
For the cumulative XLF, we first constructed the cumulative XLF  for each set of luminosity
estimates. 
Then,   we calculated the mean and the standard deviation of the number of sources in 
each bin.  
 Since the two sources of errors (photon counting/spectral
uncertainties for each source 
and  number of sources in each luminosity bin) are
independent, we combined them in quadrature. 
The differential and cumulative XLFs 
 calculated as described above are presented in the  bottom left and right pannels of  Fig. 4 respectively. 

 A comparison between the  XLFs shown in the top and bottom pannels of Fig.~4 indicates that
they are slightly different: a few sources in the second XLF have
luminosities below our completeness limit and a hump present in the
first XLF disappears in the second. This is mainly 
 an effect of the intensity-dependent spectral models used to estimate
the luminosities of the  
 sources (Table~1), which causes  lower luminosity sources
to occupy lower luminosity bins than  for the
 5~keV Bremsstrahlung spectral model used in
\S2.2. From these 
figures is also clear that while the two XLFs have similar 
 errors in the high luminosity bins (where errors are dominated by source
counting statistics), in the lower luminosity bins the
uncertainties of the second XLF are larger since the contibution of
photon counting and spectral uncertainties is becoming important for
fainter sources.

\subsection{Fits to the XLF}

We perform a Maximum Likelihood fit to the unbinned differential XLF
following the method of   Murdoch \etal (1973), which takes into
account mesurement errors. In order to be able to directly compare our
results with results on other galaxies, we only include errors
associated with the photon statistics for which we adopt the Gehrels
approximation (Gehrels, 1986). We fit the XLF with a power-law
of the form
$\rm{\frac{dN}{dS}=A\times S^{-\alpha}}$, where N is the number of
sources and S is the number of counts for each source. For the 
exposure of this observation and  a 5~keV thermal Bremsstrahlung model
with  Galactic
line of sight  absorption, 1~count corresponds to an absorption
corrected luminosity of $10^{37}$~\ergs.  
We find a best fit slope of  $-1.52^{+0.08}_{-0.33}$ (the errors are
at the 90\% confidence level; $\Delta Log(L)=1.36$, where L is the
likelihood function). Then from the number of detected sources 
with luminosities above our detection limit
(12 counts or $1.2\times10^{38}$~\ergs),  we determine the
normalization of the differential XLF  to be $22.29^{+1.39}_{-0.32}$.
To investigate the effect of the  errors on this fit we also employed
the method of Crawford \etal (1970) which is also a Maximum Likelihood
 method but it asuumes no measurement errors. In this case we obtain
very similar results but with smaller errors ($\alpha=1.57\pm0.12$).

As a check of the above results we also fit
 the binned cumulative XLFs with the \sherpa fitting package (Freeman
\etal 2001), which is part of the CXC-CIAO tool suite. We assumed a
power-law of the form  
$\rm{N(>L_{X})=B(\frac{L_{X}}{10^{38}ers/s})^{-\beta}}$, where
$\beta=\alpha - 1$  and we  used
the Cash statistic (Cash 1979), because in the highest
luminosity bins there are very few sources and the errors cannot be
described by Gaussian statistics. 
 Using  the XLF of Fig~4b which includes only source counting errors  gives  a slope of
$\beta=0.47^{+0.05}_{-0.05}$ and a normalization 
$\rm{B=46.12^{+4.4}_{-4.0}}$.
The cumulative XLF together with the best fit model and the residuals
after the fit are shown in Fig. 4b. The errors are at the 90\% confidence
level; $\Delta Log(L)=1.36$, where L is the likelihood function.
 The shape of the XLF suggests that there
may be a break around $10^{39}$\ergs. However, fits with a broken
power-law could not be justified by the existing data, since the
improvement in the fit is not statistically significant.

A fit of the cumulative  XLF which also includes errors due to photon and
spectral uncertainties (Fig.~4d)
 with the same model 
gives very similar results  and the best fit slope is
$0.45^{+0.05}_{-0.05}$. The normalization is
$\rm{B=38.9^{+3.9}_{-3.8}}$.   The best fit model and the resulting
residuals are shown in Fig. 4d. 
Since this is the most accurate determination of the parameters
of the XLF, we will use these results in our analysis.

Although we find that the two methods for 
the determination of the errors  in the XLF
 give very similar results, for XLFs with larger number of sources, 
the standard Poissonian (or Gehrels) errors may underestimate 
 the real uncertainties, especially for the lowest luminosity
 bins. This is because given the shape of the XLF, there are  
larger numbers of sources in the lower luminosity bins 
 and therefore smaller errors if only source counting uncertainties are assumed. 
 However, fainter sources have larger photon counting and spectral 
uncertainties  which may well dominate over the 
source counting errors. Therefore both sources of uncertainty should
be taken into account in the derivation of the XLF.

\section{Discussion}

In the previous section we have derived the XLF of the discrete sources in the Antennae galaxies.
Now we will use these results, together with the spectral and multiwavelength properties 
of the sources presented in Paper~III, in order to constrain the nature
of these X-ray sources.
 We will focus on the enigmatic ULX source population
($\rm{L_{X}>10^{39}}$\ergs, well above the Eddington luminosity of
a spherically accreting neutron star) which is uniquely rich in the Antennae.
A discussion on the nature of the lower luminosity sources was
presented in Paper~III, where we concluded that they were likely to be
similar to well-known Galactic sources.

\subsection{Luminosity function}

The shape of the Luminosity Function of the X-ray source population in
a galaxy can provide constraints on the nature of the sources. For
example, theoretical modeling by Wu (2000) suggested that the aging of
the X-ray source population and successive bursts of star-formation
 may produce discontinuities in the XLF. 
Constraints  on the evolution and the nature
 of the X-ray sources can also be set by comparing the 
 XLFs of galaxies in various evolutionary
stages.
The cumulative XLF of the Antennae galaxies is well fit with a single power-law
with a slope of  $\alpha=-0.45$.
This is much flatter than that
found in the bulge of M31 ($\alpha=-0.8$ Trinchieri \& Fabbiano 1991) or other early type
galaxies in the same luminosity range
 (e.g. NGC4697 where $\alpha=-0.29$ for \lx~$<3.2\times10^{38}$\ergs~ and
-1.76 for \lx~$> 3.2\times10^{38}$\ergs; Sarazin \etal 2000),
for the same formalism of the XLF.
However, the measured slope is similar to the XLF of M82
(Zezas \etal 2001) and  of the disk population of M81
 ($\alpha=-0.5$) (Tennant \etal 2001).  It is also similar to the
slope of the HMXB population of our Galaxy as measured by Grimm \etal (2001).
A comparison between the XLF of the Antennae   with that of other
galaxies is given in Fig. 5, where we see that the XLFs of galaxies with some level of
star-formation  do not show  the break reported  in
early type galaxies.   In Figure 6 we plot the logarithm of the
normalization of the XLF (at $10^{38}$\ergs) against its slope for
star-forming and early-type galaxies (for the early type galaxies we
use the slope of the high end of the XLFs,
$\rm{L_{X}>10^{38}}$\ergs).  In this plot we clearly see that
star-forming galaxies are well separated from more evolved
systems. The difference in the normalizations is most probably 
related to the stellar content of the systems (although as pointed out
by Grimm \etal (2001)  the XLF of a galaxy is
not a secure measure of its star-formation rate).
  The same effect is seen when we
compare the slopes of the XLFs with the U-B colours and 
$\rm{H_I}$ surface density of the galaxies. The former gives a measure of recent
star-formation, since is very sensitive on the number of early type
stars. However, the U-B index is also sensitive to obscuration which
 makes it very difficult to use as an accurate star-formation
indicator. The $\rm{H_I}$ surface density
instead is directly related to the surface density of the star-formation
rate by the Schnidt law  (e.g. Kennicutt, 1998). This star-formation
indicator also suggests that more actively star-forming galaxies have
 flatter XLFs than early type galaxies.     A
more thorough investigation of this effect including galaxies with
intermediate activity levels is presented in Kilgard
\etal (2001, in prep.; see also Prestwich, 2001). 

The difference between the XLF of early-type galaxies and
star-forming galaxies appears to be related to their stellar populations. 
X-ray sources  in E and S0 galaxies as well as in  bulges of
spiral galaxies (eg M31, M81) are likely to be XRBs
 with a low mass (later than G type) companion  (LMXBs), since this is
the dominant stellar population in early type galaxies.
These systems become active after the donor star fills its Roche lobe,
an event which takes place $\sim0.1-1$~Gyr after the formation of the
binary (eg Verbunt \& van den Heuvel 1995).
Instead, in star-bursting galaxies,  of which the Antennae are
an  example, the X-ray source population, while including an
older LMXB component, is likely to be dominated by luminous 'young'
short lived
sources:  X-ray binaries with O and B type
companions (HMXBs) and luminous 
SNRs, since early type stars
 are the dominant stellar population.  

 In the Milky Way, SNRs and most HMXBs are not observed at
luminosities above $10^{38}$\ergs~ (e.g. Grimm \etal 2001). In the Antennae instead, and to a
lesser degree in M81 and M82 (which host  lower level  star-formation
activity)  
we observe a large number of ULXs (Fig.~5). 
This excess holds even if we use luminosities calculated for
$\rm{H_{o}=75}$~\kmsmpc~ (dashed line in Fig.~5).
  The increasing numbers of ULXs with increasing starburst activity
associates  these sources
 with a young stellar population. If the break in the XLF of early
type galaxies is the reflection of the mass of the accretors in the
XRB population (Sarazin \etal 2000), the flat XLF of star-forming
galaxies might suggest a radically different accretor mass function or
XRBs with significantly different emission properties. 

  In what follows we discuss  possibilities for the nature of
 the ULXs, in the light of the XLF and their spectral and multiwavelength properties 
 presented in Paper~III.

\subsection{ The nature of the ULXs}

 Extranuclear  ULXs with luminosities in excess of
$10^{39}$\ergs~ and as high as
$\sim10^{40}$\ergs, were known from
previous studies with \einstein, ROSAT and ASCA  (e.g. Fabbiano \etal 1995;
Roberts \& Warwick, 2000, Colbert \& Mushotzky 1999, Makishima \etal 2000). The most extreme
example is the source recently discovered in the central region of M82,
which reaches luminosities higher than $10^{40}$\ergs~  (Kaaret \etal 2001,
Matsumoto \etal 2001).  ULXs are known to vary  and tend to have hard spectra which can be well
modeled with a  
multi-temperature disk black body model 
(disk-BB) with a temperature of $\sim1-2$~keV (e.g. Zezas \etal
1999; Makishima \etal 2000). Also, some of them 
experience hard/low - soft/high transitions typical of black hole candidate binaries
(Kubota \etal
2001; La Parola \etal 2001).  As discussed in Paper~III, 3 of the ULXs in the
Antennae are variable and their X-ray spectra are consistent with the
models described above. 

Two scenarios have been advanced to explain the high luminosities of
ULXs: the first,  which has been widely 
discussed in the literature, is that of accretion binaries with
 a BH of mass in the range $10-1000~\rm{M_{\odot}}$ 
(intermediate mass black-holes; IMBHs) (e.g. Fabbiano \etal 1995;
Marston \etal 1995; 
 Ptak \& Griffiths 1999; Matsumoto  
\etal 1999;2001; Zezas \etal 1999; Makishima \etal 2000; Kubota
\etal 2001).  The second model, is that of a binary system with
beamed emission and does not require high mass accretors (Reynolds
\etal 1997; King \etal
2001).   Below we will discuss how these models measure against the
observations of the ULXs in the Antennae galaxies.

\subsubsection{ Intermediate Mass Black holes (IMBHs)}

 Because of the high disk-BB temperatures suggested by the data, in IMBH  
models ULXs are explained by either accretion onto rotating BHs (Makishima et 
al 2000) or by the presence of `slim' accretion disks around the BH
(Watarai \etal 2000).  Formation scenarios for the ULXs include first
the formation of 
the IMBH, either by  merging of stars in a cluster 
(Portegies Zwart \etal 1999) followed by direct collapse into a IMBH (Arnett 
1996), formation of a massive object in the core of a dense cluster
(Portgies Zwart \& McMillan 2002), by merging of smaller BHs in a
cluster (Taniguchi et al 2000), or 
as the result of the evolution of primordial population-III stars (Madau \& 
Rees 2001). Once the IMBH exists, the ULX may form by capture of a normal star 
(Fabian et al 1975, Quinlan \& Shapiro 1990, Madau \& Rees, 2001).

While any of these scenarios may explain an isolated instance of ULX, some of their
predictions  are 
hard to reconcile with the properties of the ULX population of the Antennae 
galaxies. In what follows, we discuss these constraints for each model:

\smallskip
\noindent
(1) The Portegies Zwart et al (1999) scenario would 
produce IMBHs via stellar collisions (and then ULXs by capture) in young stellar
clusters. However, the 
position of most of the Antennae ULXs is near but not coincident with such 
clusters (typical offsets are $\sim100-300$~pc; Paper III). As
discussed in Paper III, this lack of positional  
coincidence may suggest that the ULXs are runaway binaries.
This hypothesis is explored further in \S 3.2.3, where we derive
mass constraints for such a system that exclude IMBHs.

Even if they receive kicks,
massive remnants  would be  expected to return to their parent cluster within a few
crossing times (Portegies Zwart \etal 1999). This timescale is of  the order
$10^{5}$~yr for the cluster R136, the dominant cluster in 30Dor in the
LMC.
The stellar clusters in the Antennae are at least 10 times more
massive than R136 (Whitmore \etal 1999), resulting in an even  shorter return timescale
($t \propto M^{-1/2} r^{3/2}$, where $M$ is the mass of the
cluster and $r$ is its half-mass radius; Portegies Zwart \etal 1999). 
Therefore, one expects only to find massive objects in stellar
clusters or close to the nuclear regions of galaxies, in contrast to
what we see in the Antennae.

\smallskip
\noindent
(2) As mentioned by Portegies Zwart \& McMillan (2002), isolated IMBHs  
may be the remnants of evolved (and therefore dissolved) dense star
clusters.  However, the   typical masses for the young star-clusters in the Antennae
($10^{4}-10^{6}$~\msun; Zhang  
\etal, 2001) imply relaxation timescales which would  prevent
 core collapse and formation of IMBHs
(Portegies Zwart \& McMillan  
2002).

\smallskip
\noindent
(3)
In the Taniguchi et al (2000) model, the IMBHs formation
timescale is $\sim 1$~Gyr. In this model, the IMBHs would be likely to be 
found in  globular clusters or galactic
nuclei. Clearly this model does not predict the preferential association of
ULXs with actively star-forming galaxies, shown by the comparison
of galaxian XLFs (\S 3.1.). 

\smallskip
\noindent
(4)
In the Madau \& Rees (2001) model, the IMBHs would eventually drift into 
globular clusters or nuclei because of dynamical friction. However, the 
time scale for this process is long enough ($\sim10^{3}$~Gyr for
a 150\msun~ black-hole, 1~kpc from the nucleus) 
that we could observe some of these 
IMBHs while they drift through the galaxy after having captured a companion 
star in a dense star cluster. The capture timescale is proportional to
the stellar density  of the cluster 
\footnote{The rate for such captures is proportional to the stellar
density: $ r = 2\pi 3.15\times10^{-24}
(\frac{\chi}{\rm{R_{*}}})(\frac{R_{*}}{\rm{R_{\odot}}})(\frac{\rho}{\rm{pc^{-3}
}})(\frac{M + M_{*}}{\rm{M_{\odot}}})(\frac{v_{d}}{\rm{km/s}})^{-1}  \rm{s^{-1}}$
(Fabian \etal 1975, Quinlan \& Shapiro 1990), where 
 $M_{*}$ and $R_{*}$ are the mass
and the  radius of the
star respectively, $\rho$ is the density of the cluster, $\chi$ is the capture 
radius, $u_{d}^{2}$ is the stellar velocity
 dispersion and M is the mass of the compact object. },
 hence the
required density for a capture  timescale of  50~Myr (the typical age
of  the optical counterparts of the ULXs in the Antennae) is
$\sim3\times10^{5}~\rm{pc^{-3}}$. Indeed,
in some star clusters in starburst regions 
such high densities can be found (e.g. Massey \& Hunter 1998).  
So, at this moment, this model is still a 
possibility for the ULXs in star-forming galaxies.

\medskip
Both the Taniguchi and the Madau \& Rees models may be relevant for ULXs
in E and S0  
galaxies, where these
sources may be associated with globular clusters (e.g. Angelini et al 2001).
The relative paucity of ULXs in early-type galaxies (see Fig.~5)
cannot be due to the effect of dynamical friction, since the
timescales are much larger than the ages of star-clusters in these
galaxies and  the high stellar densities 
would  foster the formation of many LMXB-IMBH systems. 
 It is also unlikely that it may be connected with short life-times for these 
sources. Even if accretion rates scale with the mass of the IMBH, resulting in 
100-times higher rates than in LMXB-BH Galactic binaries, we would still 
expect these binaries to live $10^{5}-10^{6}$~yr after they enter
their X-ray emitting  
phase, which begins $\sim10^{8}$~yr after their formation.
Instead, the lack of 
ULXs in early-type galaxies may suggest that `potential' ULXs with low mass
companions  are X-ray transients with very long recurring timescales. 
Indeed, black-hole binaries with evolved companions are expected to
form soft X-ray transients (King \etal 1996).
Another possibility is that these systems went through a thermal timescale mass
transfer episode resulting in a common
envelope situation (e.g. King \etal 2001). In this case the X-ray
binary is not observable  because  the X-ray emission is
attenuated by the high density material (see next
paragraph). 

\subsubsection{ Super-Eddington accretion and beaming}

 An alternative model  for  the ULXs is
that they  are beamed X-ray binaries, with normal mass accretors
(neutron star or stellar mass BH). There are two cases of beaming
which can apply to ULXs :  relativistic beaming from a jet (e.g. Reynolds \etal 1997), and
anisotropic emission from an accretion disk in a supercritical
accretion regime (e.g. King \etal 2001).
Examples of relativistically beamed sources would be the elusive
microblazars (e.g. Mirabel \& Rodriguez 1999). In the case of
microquasars, which also exhibit relativistic jets, is unlikely that
the bulk of 
their X-ray emission is relativisticaly beamed (at least while in high
state; see Fender (2001) for the contribution of jets in the low
state), since their X-ray spectra clearly show signatures of emission
from an accretion disk (e.g. Miller \etal 2001; Makishima \etal
2000),  which otherwise would be outshined by the jet. Similarly, since the
coadded spectra of the ULXs in the Antennae suggest emission from an
accretion disk (see Paper~III)  most probably the bulk of the
X-ray emission in most of these sources is not relativistically beamed. 
Recently Kording \etal (2001) by  modeling the composite XLF of three
nearby galaxies (M31, M101, M82) with a population of  
X-ray binaries with jets, constrained the contribution of the jet to
30\% of the emission  
of the disk. This contribution is in general agreement with the disk
dominated X-ray spectra   of  microquasars and ULXs, but it requires a
jet component which  should be visible in high   quality spectra. This
could be associated with the power-law component detected in these   sources 


The relativistic beaming factor $b$ for a given Lorentz factor
$\gamma$ (and the coresponding $\beta=u/c$) is a function
of the angle $\theta$  between the jet axis and the line of sight:
$b=[\gamma(1-\beta cos\theta)]^{-p}$, where $p$ depends on the
parameters of the jet (a reasonable value is 4; e.g.  Urry \& Shafer,
1984;  Mirabel \& Rodriguez 1999). From this formula one derives 
 a maximum beaming factor is $\sim(2\gamma)^{p}$ for a pole-on view of the
jet. Assuming that the most luminous source in the Antennae is beamed
with the maximum beaming factor,  we can estimate the viewing angle for
the least luminous source detected in our observation, assuming that
both belong to the same parent population and the luminosity
difference is just an orientation effect : 
\begin{equation}
\frac{L_{max}}{L_{min}} = (\frac{b_{max}}{b}) = [2\gamma^{2}(1-\beta cos\theta)]^p
\end{equation}
  From this equation, for the measured luminosity ratio of  120,
and values of $\gamma=3$ ($\beta=0.92$, beam solid angle 0.1~sr) and
$\gamma=5$ ($\beta=0.98$, beam solid angle 0.04~sr)
(e.g  Mirabel \& Rodriguez 1999) we obtain viewing angles of 30\deg
  and 18\deg  respectively.  Assuming random
orientation of the jets on 
the sky,  the ratio between maximally beamed sources and  
sources with lower beaming factors due to
different orientation is equal to the ratio of their beam solid
angles.
This predicts  8 sources, down to our limiting
luminosity ($\sim10^{38}$\ergs), for each beamed
source for both values of $\gamma$.
 Therefore for the 4 sources above $10^{40}$\ergs, we expect to
have 32 sources down to our detection limit. This is more than 75\% of
the total observed population and may agree with the model of Kording
\etal (2001).

 Anisotropic emission is a natural consequence of thick accretion
disks (e.g. Abramowitz \etal 1980, Madau 1998), which are consistent
with the X-ray spectra of ULXs  (e.g. Watarai, \etal 2000; Kubota
\etal 2001).  In this case
the luminosity enhancement is not  due to relativistic effects
(Lorentz boosting) as in the case of jet  dominated sources, but to
emission from the inside walls of the funnel, which is believed to
form in the inner part of thick accretion disks (e.g. Abramowicz \etal
1980).
This beaming was invoked by King \etal (2001)
in order to explain the large number and high observed luminosities of ULXs
in the Antennae.  In this
model the luminosity calculated under the assumption of isotropic
emission is overestimated by a factor $b$ which depends on the viewing
angle and the physical parameters of the accretion disk (e.g. Madau,
1988; Urry \etal 1991).

 King \etal (2001) proposed that one class of objects which may go
through a phase 
 of super Eddington accretion are close X-ray binaries  experiencing
 a thermal timescale 
 mass transfer episode: the atmosphere of the donor
expands,  over-filling its Roche lobe, and large amounts 
 of material flow towards the compact companion, resulting in super-critical accretion. 
 These episodes can be experienced by relatively early type 
stars  when they  evolve off the Main Sequence. However, in order to
produce an observable XRB, the  donor should not have a convective envelope (i.e. it
should be earlier than A-type for a neutron star accretor) 
(A. King, private communication).
 Otherwise,
the mass transfer episode will give rise to a common envelope binary. 
Although models show that
super-Eddington accretion can also occur in common envelope X-ray
binaries (Chevalier 1993, 1996; Fryer \etal 1996), the very large
optical depths of the convective stellar atmospheres will make any
produced X-ray emission unobservable, since it will be attenuated by
absorption as well as  scattering.

Recently Pacull \& Mirioni (2002), based on the detection of high
excitation lines in the optical spectrum  of the ULX in the nearby
galaxy Ho-II, suggest that the local ISM sees an ionizing continuum of
$\sim10^{40}$\ergs and hence the X-ray emission cannot be beamed.
However, it is possible that highly ionized nebulae can also be
produced by beamed X-ray  binaries with the only difference that they
would have a bi-conic  morphology  (similar to that of the ionization
cones in Seyfert galaxies) if seen edge on. If seen face-on they
should be indistinguishable from any nebulae ionized by an isotropic
source.

\subsubsection{`Runaway' constraints to binary masses}

In the beamed model, the positional offsets of the ULXs from the
stellar clusters (Paper~III) is explained easily by the runaway
binary scenario.
For beaming factors as low as 0.1, the mass of the compact object
reduces to 1-10~\msun. Objects of this mass can easily receive large
kicks when their progenitor undergoes supernova explosion 
($\rm{v\sim100-30} $~\kms~ respectively; Cordes \& Chernoff 1998; Fryer \&
Kalogera, 2001) and be expelled from their parent clusters.
The presence of these displacements, in the assumption that
they are related to kicks, can help us further constrain the nature of
the binary system.

 The final velocity of the system depends both on the mass of the
compact object and of the companion star. 
Rescaling from the Cordes \& Chernoff (1998) kick
velocities for neutron stars,  we obtain
 $v = 245*(M_{BH} + M_{comp})^{-1}~\rm{km~s^{-1}}$, where
$M_{BH}$ is the mass of the compact object and $M_{comp}$ is the mass
of the companion. In Fig. 7 we plot the timescale needed by the system to
cover a distance of $\sim300$~pc (which is the typical offset between
the X-ray sources and their nearest optical counterparts; Paper~III)
as a function of the mass of the donor, for different masses for the
compact object. Note that as discussed above, later than A type donors
are excluded (hatched area). This is a limiting case for neutron star
accretors; if the accretor is more massive then this limit moves to
higher mass donors in order to avoid the formation of a transient source (King
\etal 1996). Most high mass X-ray binaries enter their X-ray
emitting phase soon after their donors evolve off the Main Sequence (e.g. 
Verbunt \& van den Heuvel, 1995), therefore if the ULXs are HMXBs they
should have been expelled from their parent clusters in a timescale
comparable  to the companion's Main Sequence lifetime (Fig. 7). 
Imposing that the age of
the companion should be at least similar to
 the timescale required by the system to move to its present position,
results in an upper limit to the permitted donor masses
given by the abscissa of the point where the Main Sequence
lifetime- mass relation (dashed line) intercepts the solid lines in
Fig.~7.   Since the kick
velocities can be much lower than those assumed for this plot,
companions with lower masses would be consistent with the measured
distances.

If the companion is too massive,  the system would become active before it
has an appreciable separation from its parent cluster; these massive binaries
could be the counterparts of sources associated with star
clusters. Alternatively, these objects could host massive IMBHs, as
discussed in \S3.2.1. 

Fig. 7 suggests that sources with measured
offset from a nearby star cluster  (Paper~III) may be binary systems
with donor stars ranging from B2 to A type for a range of compact
accretors from a 1.4\msun~ neutron star to a 20\msun~ black-hole.
In these systems, super-Eddington
accretion will lead to the formation of a thick accretion disk, as
discussed earlier.  
This is not a problem even in the case of a magnetized neutron star
(which may be common in young binaries),  since polar accretion
can well take place in the center of a thick accretion disk (A. King
private communication).
 Moreover, the large magnetic fields may reduce
the Thompson scattering cross-section and thus increase the Eddington
limit (e.g. King \etal 2001).
In the case of a strongly magnetized neutron star, one expects
to observe strong cyclotron lines (as in pulsar binaries in our Galaxy; e.g. Nagase 1989) 
if the X-ray emission is produced close to the compact object. This
could be a way of discriminating between black hole and pulsar accretors.

 This model could also be used to connect the ULXs with the Galactic
microquasars (e.g. Mirabel \& Rodriguez 1999). As presented in
Paper~III, both types of sources have  similar parameters for the
disk black-body model used to fit their X-ray spectra, suggesting a similar
structure in their accretion disks.  In this case the factor of 10
difference in their luminosities may be explained in terms of
different beaming factors. Indeed if microquasars accrete at lower
rates than ULXs (but still super-Eddington, ), the formed accretion disk
will be slim (Abramowitz \etal 1988), in agreement with spectral
fitting results (Watarai \etal 2000). This disk lacks the large
funnel-like structure of thick disks, resulting in much lower beaming factors
 which however are still higher than in standard thin disks (e.g. Abramowitz \etal
1988). The lower accretion rates could be related to the details of
the mass transfer and the orbital parameters of the binary.
 In this context at the upper end of the ULX population in the
Antennae could be objects similar to those proposed by King \etal
(2001), while lower luminosity sources could be similar to Galactic
microquasars.  

\subsubsection{Effect of beaming on the XLF}

 Beaming has  important implications for the observed XLF of the
ULX  population. Urry \etal (1991; see also Celotti \etal 1993)
studied this effect in AGNs  based on a calculation of the beaming factors for
thick accretion disks orbiting supermassive black holes
(M$\sim10^{8}$~M$_{\odot}$; Madau 1988). Their results apply to the
XLF of the Antennae if  the properties of the accretion disk
(opening angle and size of the funnel, temperature structure and size of the
disk) do not depend on the mass of the central object.  
They found  that for a power-law parent
luminosity function between luminosities $l_{1}$ and  $l_{2}$,
 the beamed XLF is flatter than the parent XLF from a luminosity
$L_{1}$ (which is the minimum parent luminosity times the minimum
beaming factor), up to a luminosity of
$L_{3}=Acos(\theta_{o})\times l_{1}$, where $\theta_{o}$ is the opening
angle of the funnel.
At higher luminosities the beamed XLF follows the same slope 
 as the parent XLF (Fig. 8).  The flattening of the XLF is due to absorption of the intrinsic (unbeamed) 
 emission by the walls of the funnel when the line of sight is larger than $\theta_{o}$ (i.e. crosses the funnel).
 This behaviour is different than what one would expect for relativistic beaming where the intrinsic luminosity
 is just enhanced by a  beaming factor.
 Since the slope of the parent XLF of the ULXs actually
represents the mass function of the accreting objects,  
if they accrete at the same fraction of their Eddington rate,
 it could be used to constrain models for their formation. 

Apart from the population of ULXs, a galaxy would have a  population 
of normal XRBs (which may also have a power-law XLF; e.g. Sarazin
\etal 2000; Tennant \etal 2001). Therefore,  the total XLF for a galaxy 
 would start al low $\rm{L_{X}}$ with a power-law with the slope of the underlying XRB
population XLF and then flatten when the lowest luminosity ULXs
appear, to finally assume the slope of the unbeamed population of the
ULXs (dashed line in Fig. 8) at the critical luminosity which corresponds to the opening angle
where the emission ``funnel'' can be observed.  On the other hand 
if the Kording \etal (2001) model is valid, with X-ray binaries and
ULXs having both beamed jets and unbeamed disk components such breaks
would not be expected. Future more sensitive
observations of the Antennae will extend the range of detected sources
well into the normal XRB regime and will allow us to test different
scenaria.

From the maximum luminosity of the sources ($\sim2\times10^{40}$~\ergs)
and the maximum beaming factor predicted by Madau (1988)
 we can  set an upper limit of
$\sim10$~\msun~ on the maximum mass of the compact objects.  This limit
is fully consistent with 
stellar BHs. Since the mass of the compact object is $\sim10$~\msun~ we
are in the regime of narrow mass distribution  (case 2 of Urry \etal 1991).
In this case the highest luminosity break is expected to occur  at 
 $\sim3\times10^{39}$\ergs, if the lowest mass compact object is a
1~\msun~ neutron star and assuming critical accretion.
 The shape of the XLF of the Antennae (figs
4,5)  may be consistent with  an excess of sources below
$\sim3\times10^{39}$\ergs. In this model the first break should occur at
$\sim8\times10^{37}$~\ergs, just below our detection limit.
The flattening of the XLF of ULXs between $10^{38}$\ergs~ and
$10^{39}$\ergs~ may mask a possible break in the XLF of the normal
binaries, due to the neutron star - black hole transition. This break should
be close to  the Eddington luminosity for a neutron star (e.g.,
Sarazin \etal 2000), and existing data suggest that is not present in
star-forming galaxies (see Fig.~5).
Therefore, XLFs extending to lower luminosities will allow us to test this
model and constrain the low mass cutoff for the compact objects in the
ULXs.

\section{Conclusions}

 We have presented a study of the X-ray source population in the 
Antennae galaxies.
We derive their luminosity function by taking into account effects
related to the varying detection limit over the area of the galaxy and
including count-rate and spectral uncertainties
for each source.  We find that the cumulative XLF  has a slope
($\alpha=-0.45$), similar to that of  the 
luminosity function of the star-forming galaxy M82 (Zezas \etal 2001)
and the disk of  the spiral M81 (Tennant \etal 2001).
The XLF of the Antennae is instead much flatter than  the XLF of early type galaxies  and does not show a break around $3\times10^{38}$~\ergs~
as reported in the latter (e.g. Sarazin \etal 2001).  
These comparisons suggest that
 the dominant X-ray population in the Antennae is young.
This conclusion agrees with the results of Paper~III showing that
almost all the optical counterparts of the X-ray sources in the
Antennae are 
young stellar clusters and most of the radio counterparts have steep
radio spectra, implying SNRs in the region.
The comparison between the  XLFs of different galaxies also indicates that 
more actively star-forming galaxies have larger numbers of ULXs,
further  associating these
sources with star-formation activity and a young stellar population. 

Our results suggest that $>10-1000$s~\msun~ IMBH cannot be the sole explanation
for all the ULXs in the  Antennae, although they may explain a few of them.
In particular, the general association of the ULXs with young star clusters (Paper III) is
hard to reconcile with the Taniguchi \etal (2000) formation scenario of IMBH, 
which requires long timescales. 
The displacement of the ULXs from neighboring young
stellar clusters (Paper~III), is hard to reconcile with the  Portegies Zwart \etal (1999)
model, but could be consistent with the Madau \& Rees (2001)
scenario, in which massive black holes may be the remnants of Population III
stars, if these black holes capture a companion while crossing a dense
star cluster. However, a more likely scenario may be that of runaway
binaries with  neutron stars or low-mass black holes  which received kicks 
upon their formation.

Since it is unlikely that the majority of ULXs are powered by accretion
onto an IMBH the only way to produce the observed luminosities is by
supercritical accretion and the resulting beaming. Supercritical
accretion (which is consistent with the X-ray spectral fits  for ULXs)
has as a natural consequence the formation of a funnel in the inner
part of the accretion disk (e.g. Abramowitz \etal 1980), resulting
in anisotropic emission (beaming; e.g. Madau 1980; King \etal
2001). Using the constraint provided by the displacement of the ULXs 
from the closest star clusters where they may have formed (Paper I.I),
we derive masses of compact objects ranging from 1.4\msun~ (neutron star) to
20\msun, and stellar companions of B3-A type. Only the few object found
within a stellar cluster are likely to have a more massive conterpart.


The type of beaming produced by supercritical accretion would give a
characteristic shape to the XLF, by producing two breaks, the
position of which depends on the luminosity range of the parent
population of the ULXs (or equivalently the mass range of the
accretor). These breaks are not related either to
the aging of the X-ray source population (e.g., Wu 2000, Kilgard \etal
2001) nor to the neutron star - black hole transition in the accretors
of normal X-ray binaries (e.g. Sarazin \etal 2000). 
The flattening of the XLF of ULXs predicted by
this model may mask any breaks associated with the normal XRB
population.
 In the XLF of the Antennae we do not detect any break, but
we are limited by the sensitivity of the present observation; 
future observations of the Antennae as well as other nearby
star-burst galaxies, which will extend  to lower luminosities,
will allow us to test this model and set further constraints on the mass
distribution of the accretors. 


\acknowledgments

 We thank the CXC DS and SDS teams for their efforts in reducing the data and 
developing the software used for the reduction (SDP) and analysis
(CIAO). We thank Andrew King, Martin Elvis, Vicky Kalogera,  Andrea Prestwich,
Phil Kaaret and Jeff McClintock for 
useful discussions on these results. 
This work was supported by NASA contract NAS~8--39073 (CXC)  and NASA
Grant NAG~5--9983.

{}

\clearpage

\makeatletter
\def\jnl@aj{AJ}
\ifx\revtex@jnl\jnl@aj\let\tablebreak=\nl\fi
\makeatother
\ptlandscape
\begin{deluxetable}{lcc}
\tabletypesize{\scriptsize}
\tablecolumns{3}
\tablewidth{0pt}
\typeout{Yo}
\tablecaption{Assumed models for the calculation of the luminosity function}
\tablehead{ 
\colhead{Luminosity range} & \colhead{Photon index}   &
\colhead{Column density}   \\
\colhead{ (1) }	&\colhead{ (2)}
&\colhead{(3)}  }
\startdata
$<3\times10^{38}$	& $3.36\pm1.5 $ & $0.28\pm0.5 $ \\
$3\times10^{38} - 10^{39}$	& $3.37\pm1.0 $ & $0.23\pm0.15 $ \\
$>10^{39}$	& $1.7\pm0.5 $ & $0.1\pm0.07 $ \\
\enddata
\end{deluxetable}

\clearpage


\begin{figure}
\caption{ The uncorrected luminosity histogram  of all the sources in
the Antennae, 
compared with the uncorrected luminosity histogram of the point-like
and unobscured sources (hatched area), which we use to derive the XLF in this paper.}
\end{figure}

\begin{figure}
\caption{ An image of the 0.3-10.0~keV diffuse background in the Antennae
galaxies as calculated by the \textit{ wavdetect} tool (see Paper~II). The detected sources are marked by their $3\sigma$ positional
ellipses. The contours correspond to levels of 0.1, 0.15 and 0.35
counts/pixel. Sources outside the second contour (0.15counts/pixel)
 formed the Low background XLF, sources between the second
and the third contour formed the Medium background XLF while sources
inside the third contour formed the High background XLF.  }
\end{figure}

\begin{figure}
\caption{ Left (a):The uncorrected (solid line) and corrected
(dash-dot line) luminosity distribution of sources
found in the three different background environments.  
 Right (b):  The
correction function for these three background levels. Triangles
correspond to the high background, squares to medium background and
stars to low background levels. The dashed line shows the 50\%
completeness limit for all curves. }
\end{figure}

\begin{figure}
\caption{Top panel: Left (a): The completeness corrected differential XLF of
all the sources
 in the Antennae.  Right (b): The cumulative XLF of the sources in the
Antennae, together with the best fit power-law (solid line) and the
residuals (in number of sources) after the fit. 
%
Bottom panel: The differential (left; c) and cumulative (right; d) XLF
of the point sources in the Antennae from a Monte Carlo 
simulation which takes into 
account errors associated with count-rate and spectral
uncertainties. Together with the cumulative XLF we show the best fit
power-law (solid line) and the residuals  (in number of sources) after the fit.}
\end{figure}

\begin{figure}
\caption{ Comparison between the luminosity functions of Antennae
(this work), M82 (Zezas \etal 2001),
the disk of M81 (Tennant \etal 2001) and NGC4697 (Sarazin \etal
2001). The dashed line shows the XLF of the Antennae for
$\rm{H_{o}=75}$~\kmsmpc.  We also plot
the XLFs of the HMXB and the LMXB population of the Galaxy (Grimm
\etal 2001).}
\end{figure}

\begin{figure}
\caption{ Comparison between the slopes of the cumulative luminosity
functions of star-forming (Antennae, M82, the disk of M81) and early
type galaxies (NGC4697, NGC1553, and the bulge of M31). The ordinate
is the logarithm of the normalization of the 
XLF at $10^{38}$\ergs. Data are from: this work (Antennae); Zezas
\etal 2001 (M82); Tennant \etal 2001 (M81); Sarazin \etal 2001 (NGC
4697); Blanton \etal 2001 (NGC1553), Primini \etal 1993 (M31).}
\end{figure}

\begin{figure}
\caption{ A plot of the time required by a binary system of
various compact object  masses (given in the end of the solid line) 
 to reach a distance of $300$~pc as a function of the donor mass (solid lines). The top and right axes give the spectral
types corresponding to each donor mass and Main Sequence lifetime. The
dashed line shows the Main Sequence lifetime - mass relation. The
permitted donor types are given by the thick part of the solid
lines. The hatched area shows the region of donors with convective
envelopes, which would not produce an observable XRB, according to the model of King \etal
(2001) (see text). }
\end{figure}

\begin{figure}
\caption{ An ilustration of the effect of beaming on the parent
differential XLF of
black holes with thick accretion disks (adopted from Urry \etal
1991). $\L_{1}$ corresponds to the minimum observed luminosity for a
thick accretion disk and  $\L_{3}$ corresponds to the position of the
break in the beamed XLF.   The normalizations of the XLFs and the
slopes of the non ULX and parent XLFs are arbitrary. }
\end{figure}

\clearpage

\setcounter{figure}{0}

\begin{figure}
\rotatebox{270}{\includegraphics[height=8.0cm]{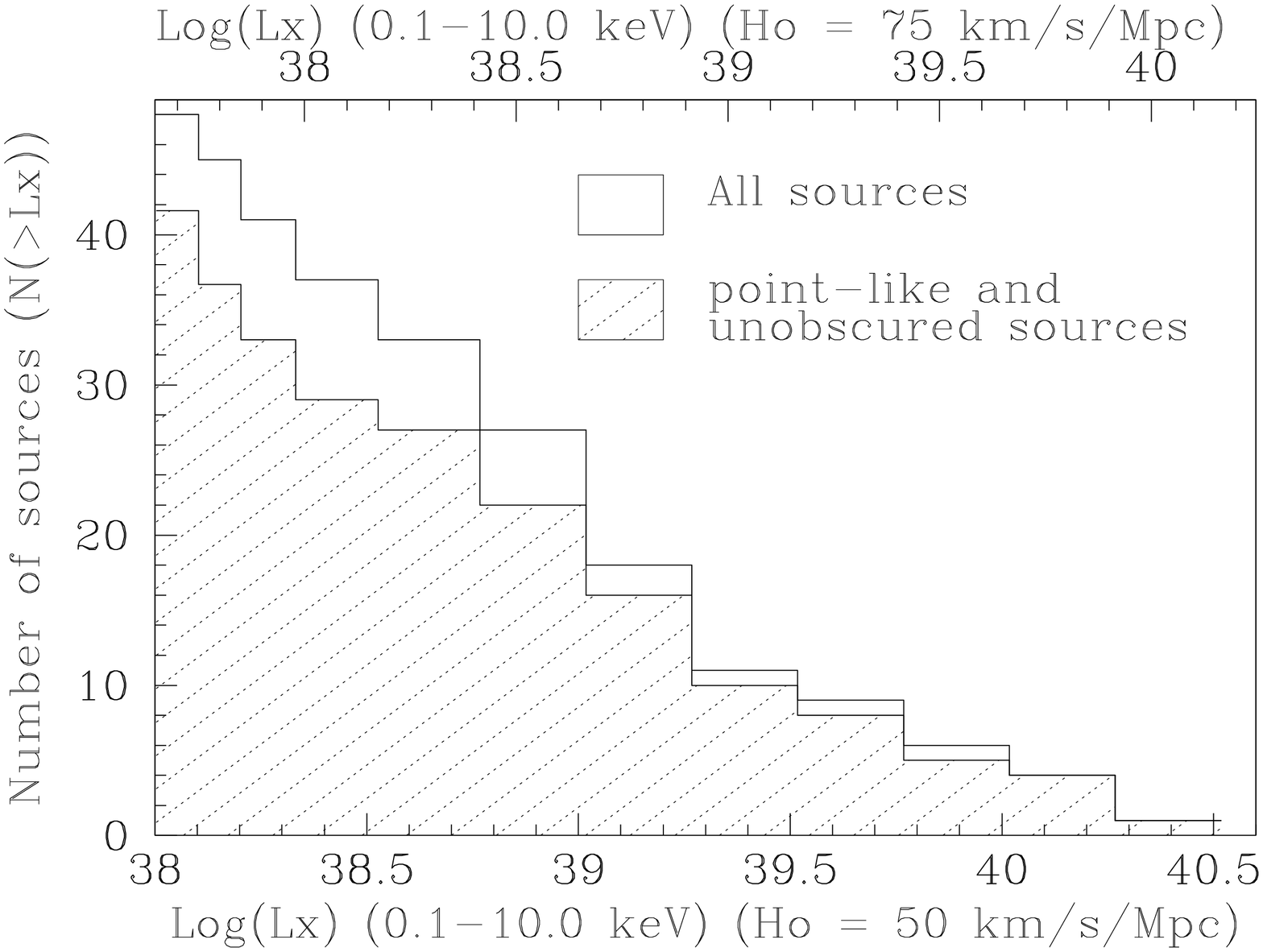}}
\caption{}
\end{figure}

\begin{figure}
\rotatebox{360}{\includegraphics[height=9.0cm]{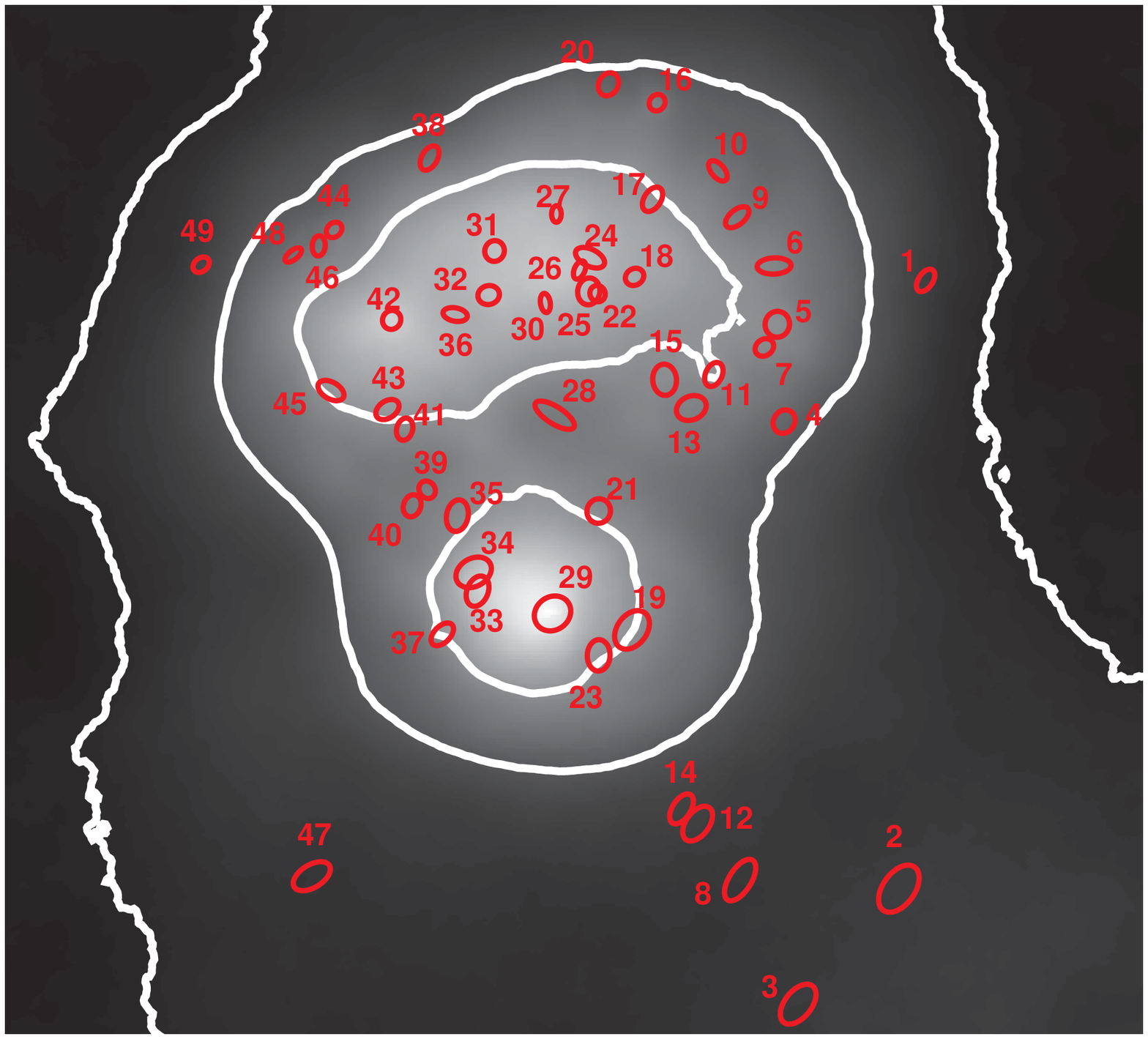}}
\caption{  }
\end{figure}

\begin{figure}
\rotatebox{270}{\includegraphics[height=7.5cm]{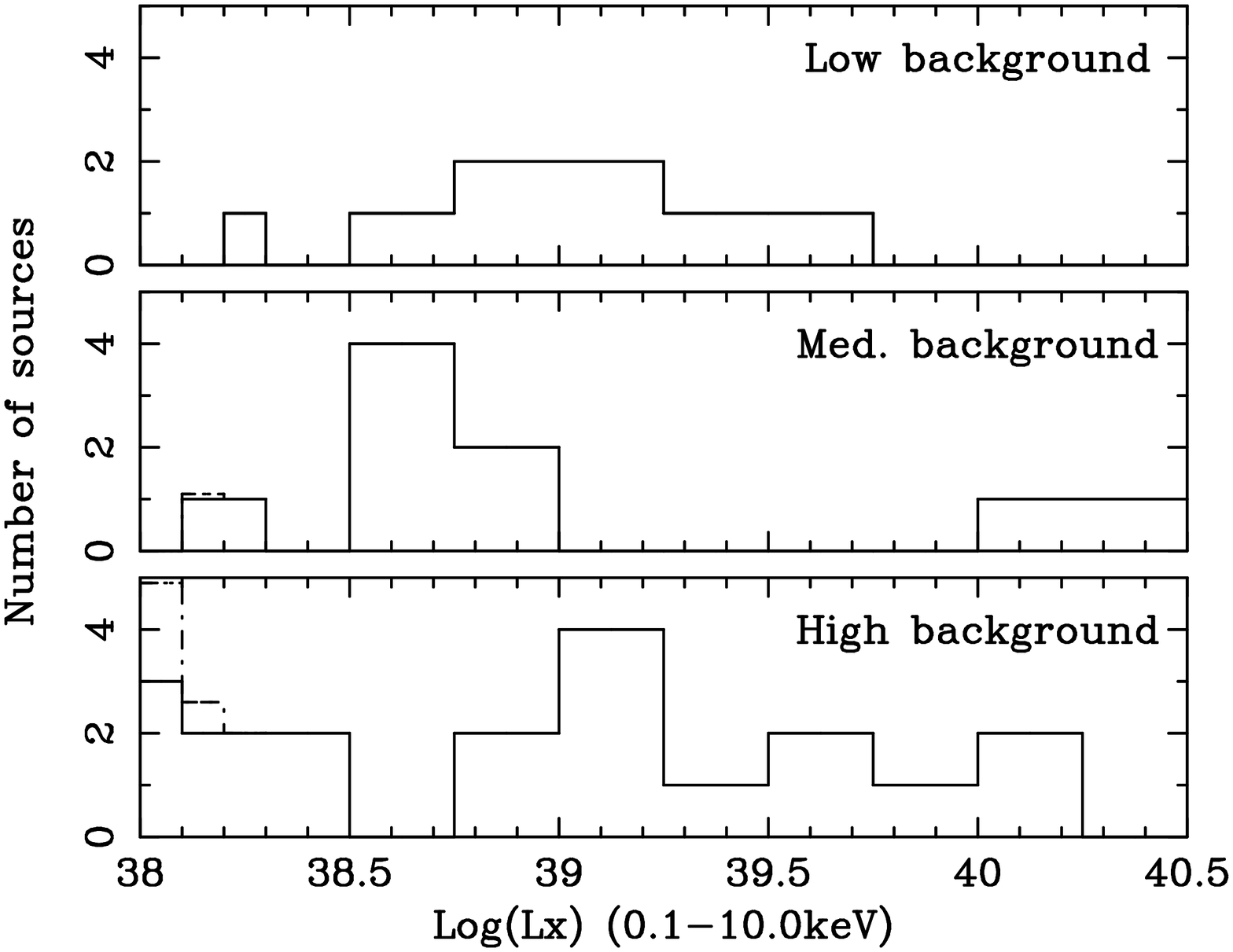}}
\hspace{0.5cm}
\rotatebox{270}{\includegraphics[height=8.5cm]{f3b.eps}}
\caption{ }
\end{figure}

\begin{figure}
\begin{tabular}{cc}
\rotatebox{270}{\includegraphics[height=8.0cm]{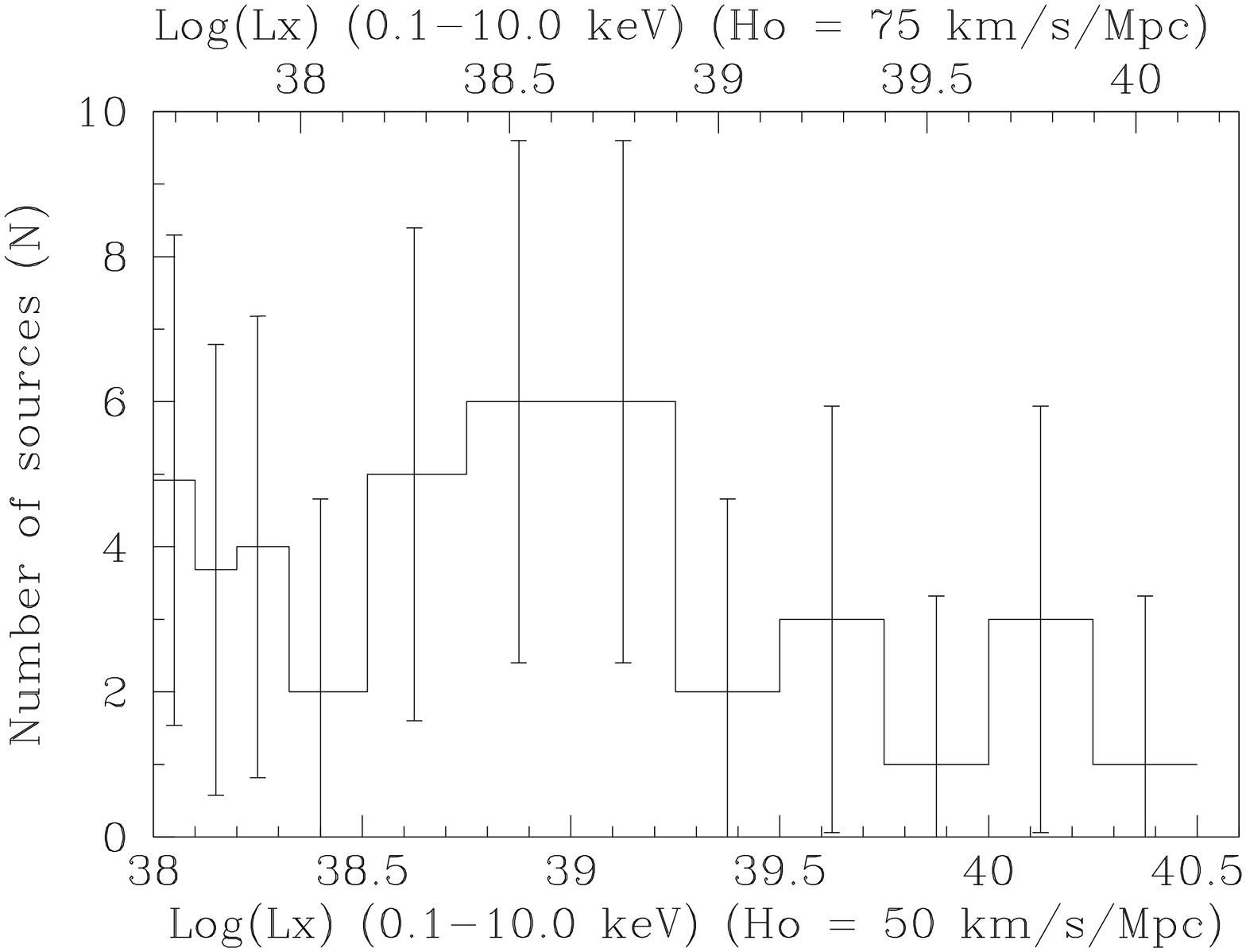}} &
\rotatebox{270}{\includegraphics[height=8.0cm]{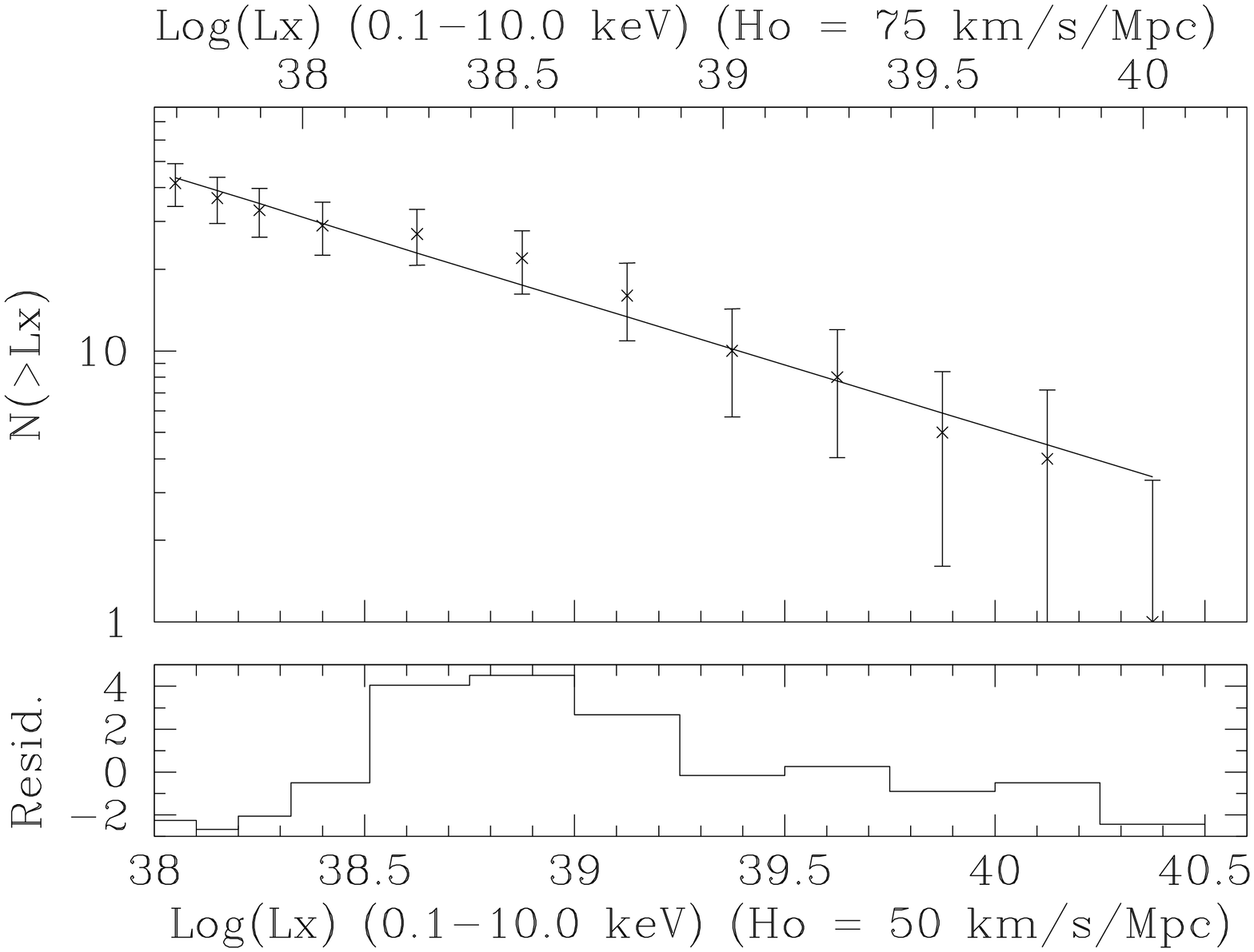}}  \\
%
\vspace{-0.5cm}
\rotatebox{270}{\includegraphics[height=8.0cm]{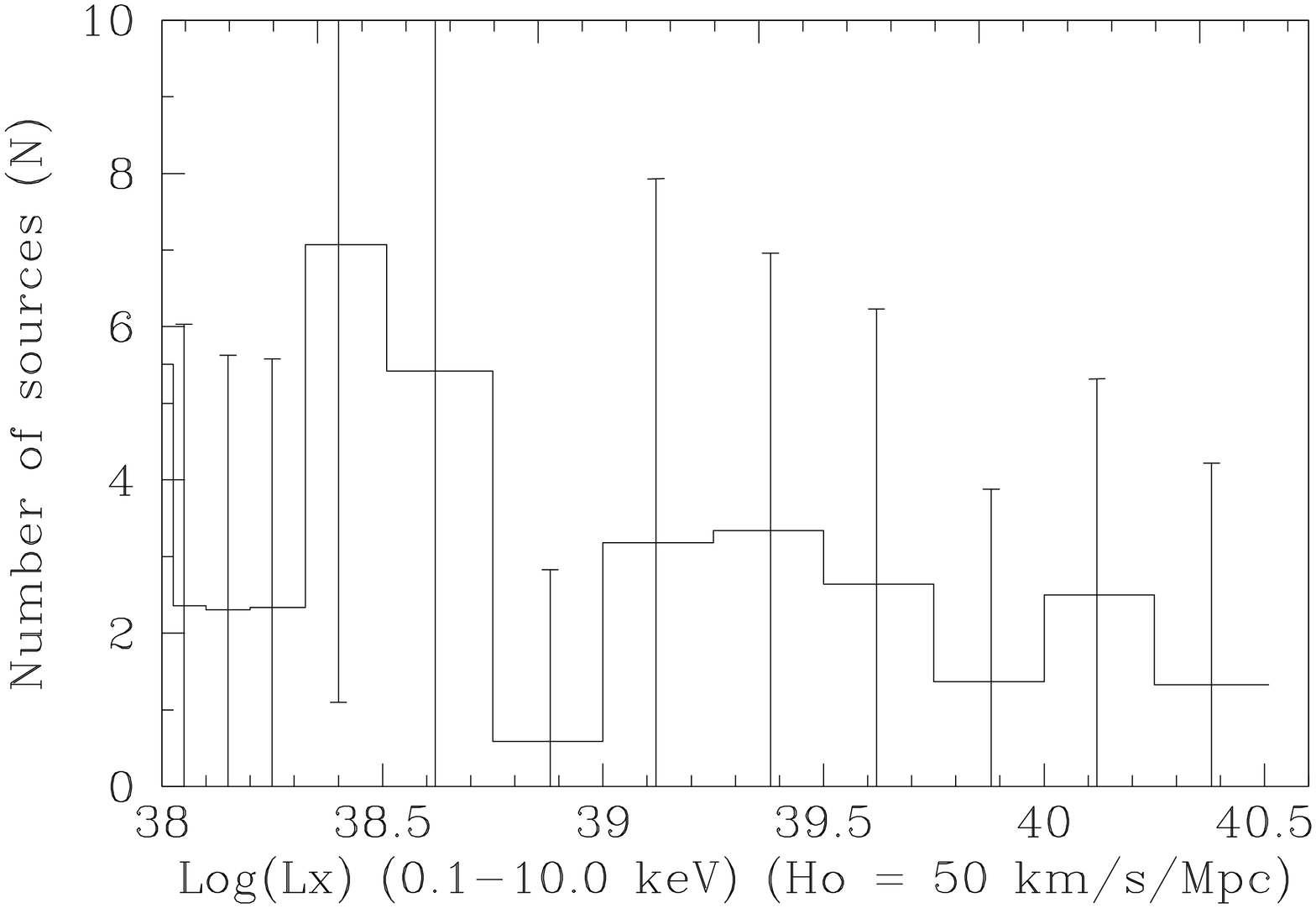}} &
\rotatebox{270}{\includegraphics[height=8.0cm]{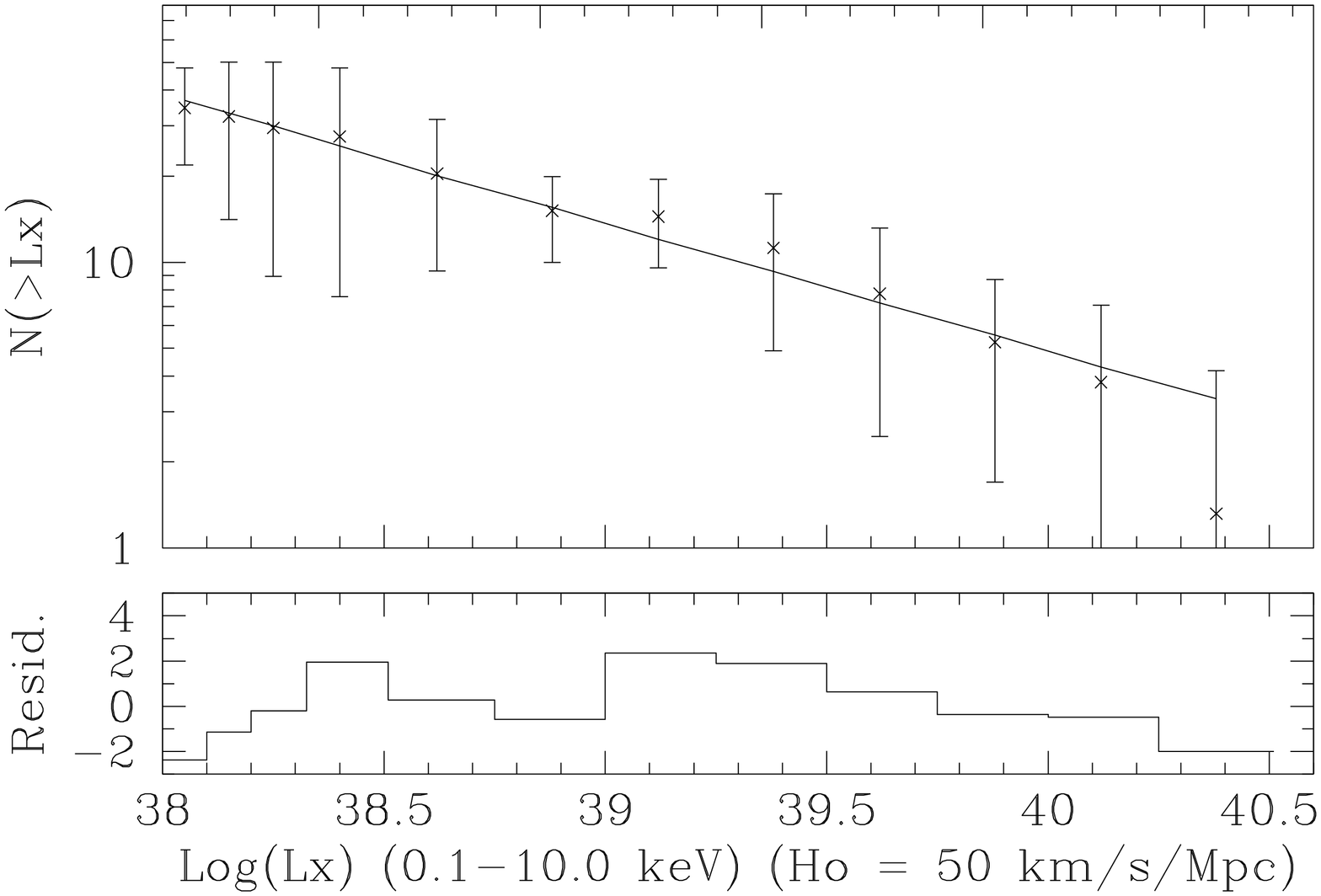}} \\
\end{tabular}
\caption{ }
\end{figure}


\begin{figure}
\rotatebox{270}{\includegraphics[height=9.0cm]{f5.eps}}
\caption{ }
\end{figure}

\begin{figure}
\rotatebox{270}{\includegraphics[height=9.0cm]{f6.eps}}
\caption{}
\end{figure}

\begin{figure}
\includegraphics[height=9.0cm]{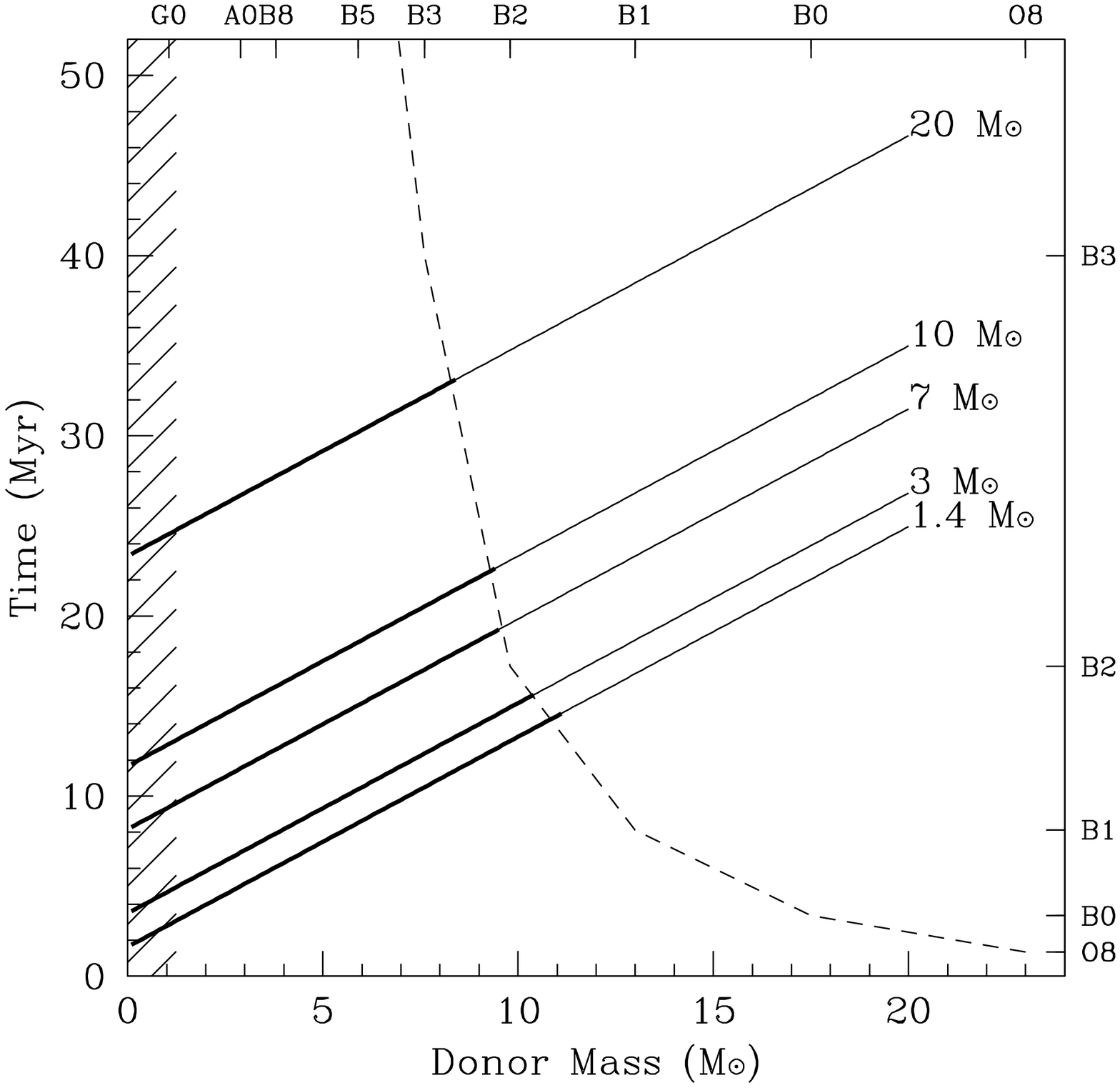}
\caption{ }
\end{figure}

\begin{figure}
\includegraphics[height=9.0cm]{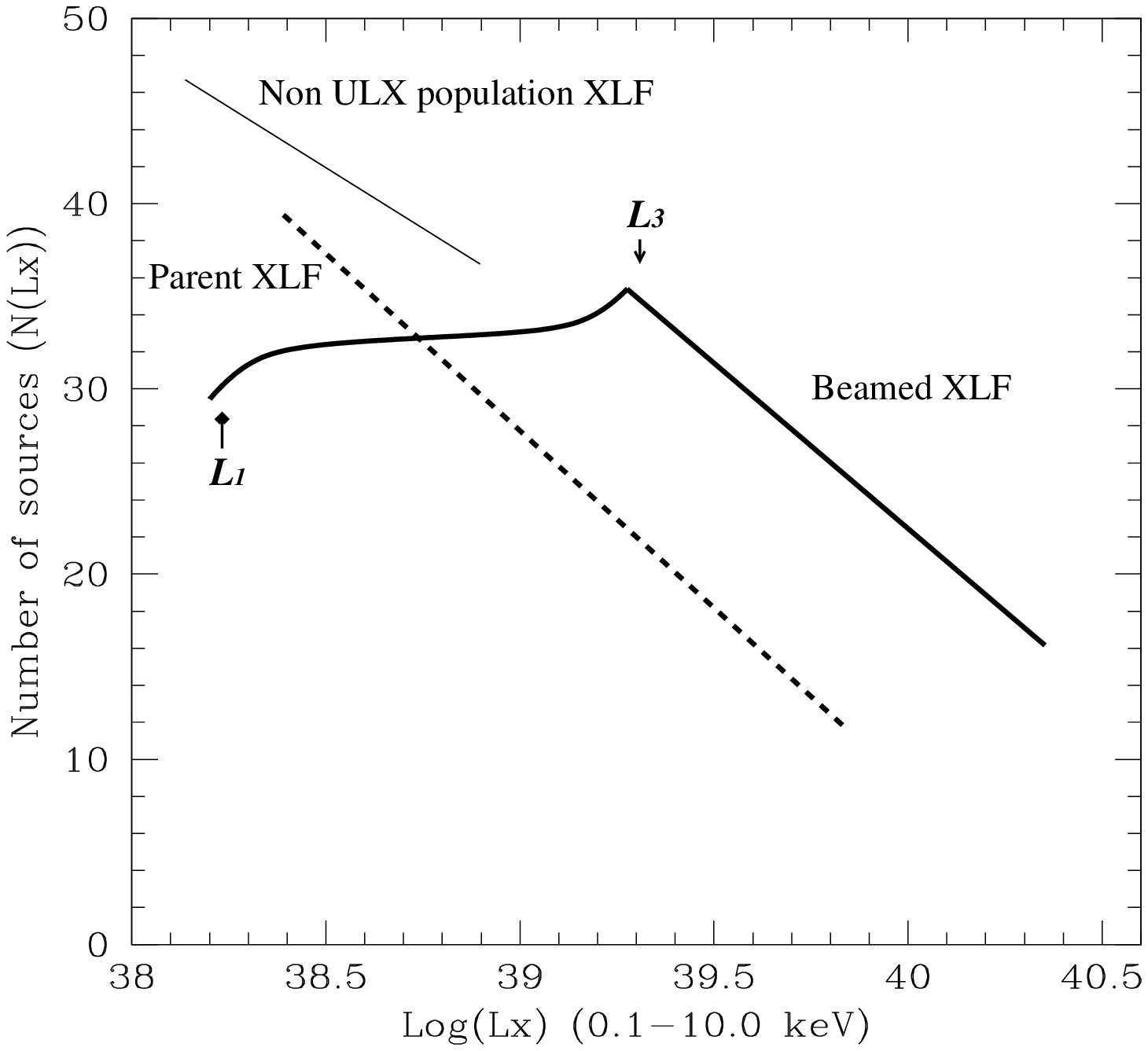}
\caption{ }
\end{figure}

\end{document}